\documentclass[epj]{svjour}
\newcommand{\eq}[1]{Eq.(\ref{#1})}

\begin{document}
\title{Vacuum polarization in muonic and antiprotonic atoms:
the fine structure at medium $Z$}
\author{E. Yu. Korzinin\inst{1},
V. G. Ivanov\inst{1,2},
\and S. G. Karshenboim\inst{1,3}\thanks{E-mail: sek@mpq.mpg.de}}
\institute{
D. I. Mendeleev Institute for Metrology (VNIIM), 190005, St. Petersburg, Russia
\and
Pulkovo Observatory, 196140, St. Petersburg, Russia
\and
Max-Planck-Institut f\"ur Quantenoptik, 85748, Garching, Germany}
\authorrunning{E. Yu. Korzinin, V. G. Ivanov, and S. G. Karshenboim}
\titlerunning{Vacuum polarization\dots}
\date{Received: date / Revised version: date}
%
\abstract{Effects of vacuum polarization modify the energy levels
in atoms with an orbiting particle heavier than an electron. The
dominant effect is due to the Uehling potential. In this paper we
consider the relativistic corrections to the energy levels caused
by the Uehling potential and in particular the fine structure in
muonic and antiprotonic atoms. We derive general expressions and
consider in detail specific regions of parameters which allow
simple asymptotic expansion. We take into account the recoil
effects and anomalous magnetic moment in the case of an antiproton
as the orbiting particle.
} 

\maketitle

\section{Introduction}

Quantum electrodynamics (QED) corrections to the energy levels of
bound states have very different structure from one atom to
another. The dominant QED effect for exotic atoms with an orbiting
particle heavier than an electron is due to the so-called Uehling
potential, which is caused by a free electron vacuum polarization
loop. In a previous paper~\cite{previous} we studied the
nonrelativistic effects which are in particular responsible for
the Lamb shift.

In principle, the results were known for a few low-lying levels
for a while~\cite{Pusto}; however, they are rather complicated and
the aim of \cite{previous} was to derive simple expressions for
arbitrary states, which may be realized as certain asymptotic
expansions. For instance, the expansions are possible over a small
parameter $Z\alpha$ (where $Z$ is the nuclear charge) and a large
parameter $Z\alpha\,m/n^2m_e$ (where $m$ is the mass of an
orbiting particle and $n$ is the principal quantum number). Some
other expansions are also possible.

Analysis in \cite{previous} was performed in the leading
non-re\-la\-ti\-vistic approximation. Here we consider the first
relativistic correction. This correction is the dominant QED
contribution to the fine structure and we pay special attention to
the fine structure effects in muonic and antiprotonic atoms.
Certain nonrelativistic asymptotics were previously studied in
\cite{CJP98,soto}, while in \cite{previous} we considered a
specific case of high $n$ which may be of particular interest due
to antiprotonic atoms.

Here we derive a general expression for the dominant correction to
the Lamb shift at a medium value of the nuclear charge $Z$. This
correction is $j$-independent  and we also found a relativistic
correction which depends on the angular momentum $j$. Finally, the
vacuum polarization corrections are presented in the form
\begin{equation}\label{defFH}
  \Delta E(nl_j)=
\]
\[
  \frac{\alpha}{\pi}\,(Z\alpha)^2\,
  \frac{m c^2}{n^2}\,\left[F_{nl}(\kappa_n)+
    (Z\alpha)^2 H_{nlj}(\kappa_n)+\dots\right]
  \;,
\end{equation}
where
\begin{equation}
  \kappa_n = \frac{\kappa}{n} = \frac{1}{n}\,\frac{Z\alpha \, m}{m_e}
  \,.\nonumber\\
\end{equation}
The first term (the $F$-term) is responsible for most of the Lamb
shift and was already studied in detail in \cite{previous}. The
relativistic correction (the $H$-term) is a small correction to
the Lamb shift, but it is responsible for the dominant radiative
contribution to the fine structure, which we define here as
splitting for the case of $j\neq j^\prime$ while $n=n^\prime$ and
$l=l^\prime$. Due to that we consider in detail the fine-structure
difference
\begin{equation}\label{deltafs}
\Delta_{nl}^{\rm FS}(\kappa_n)
=H_{n,l,l+1/2}(\kappa_n)-H_{n,l,l-1/2}(\kappa_n)\;.
\end{equation}

In the atoms of interest the mass $m$ and parameter $\kappa_n$ can
be $m\simeq 207\, m_e$ and $\kappa_n\simeq 1.5\, Z/n$ (in a muonic
atom) and $m\simeq 1836\, m_e$ and $\kappa_n\simeq13.4\, Z/n$ (in
an antiprotonic atom).

\section{The relativistic expression for the Uehling term and the
fine structure}

An exact relativistic expression for the Uehling correction to the
energy is of the form
\begin{equation}\label{reldef}
  \Delta E(nl_j) =
  \int dr \; r^2 \left( |f_{nlj}|^2 + |g_{nlj}|^2 \right) V_U(r)
  \;,
\end{equation}
where $f_{nlj}$ and $g_{nlj}$ are the upper and lower components
of the Dirac wave function \cite{IV} and for the Uehling potential
we use the Schwinger's parametrization\footnote{Throughout the
paper we apply the relativistic units in which $c=\hbar=1$.}
\cite{Schwinger}
\begin{equation} \label{Up}
V_U(r) =
 \frac{\alpha}{\pi}
 \int_0^1
 dv \,
 \frac{v^2(1-v^2/3)}{1-v^2} \,
 \left( -\frac{Z\alpha}{r} e^{-\lambda r} \right)\;,
\end{equation}
with
\begin{equation}
  \lambda = \frac{2m_e}{\sqrt{1-v^2}}
  \,.
\end{equation}

The exact calculation can be performed and analytic results for
certain states were found in \cite{CJP98}. Here, at the first
stage we derive a similar expression for an arbitrary state in a
hydrogen-like atom. Applying the known expression for the wave
function of the Dirac-Coulomb equation (see, e.g., \cite{IV}) and
integrating over coordinates, we obtain
\begin{equation}\label{eurel}
  \Delta E(nl_j)
  =
  -\frac{\alpha}{\pi}
  \,
  \eta^2
  \,
  \frac{\Gamma(2\zeta+n_r+1) (n_r)!}{\frac{Z\alpha}{\eta}-\nu}
  \times
\]
\[
  \sum_{i,k=0}^{n_r}
  \frac{(-1)^{i+k}}{i!(n_r-i)!k!(n_r-k)!}
  \,
  \frac{\Gamma(2\zeta+i+k)}{\Gamma(2\zeta+i+1) \Gamma(2\zeta+k+1)}
  \times
\]
\[
  \Biggl\{
  m\left[
    \left( \frac{Z\alpha}{\eta}-\nu \right)^2
    + (n_r-i)(n_r-k)
  \right]
\]
\[
  \phantom{mm}
  - E_{nlj} \left( \frac{Z\alpha}{\eta}-\nu \right)
  (2n_r-i-k)
  \Biggr\}
  \times
\]
\[
  \left\{
    K_{1,2,i+k+2\zeta}(\widetilde\kappa_n)
    -
    \frac{1}{3} K_{2,2,i+k+2\zeta}(\widetilde\kappa_n)
  \right\}
  \,,
\end{equation}
where
\begin{eqnarray}
\nu &=& (-1)^{j+l+1/2} (j+1/2) \,,\nonumber\\
\zeta &=& \sqrt{\nu^2 -(Z\alpha)^2} \,,\nonumber\\
\eta &=& \sqrt{1 - (E_{nlj}/m)^2} \,,\nonumber\\
n_r &=& n - |\nu| \,,\nonumber\\
\widetilde\kappa_n&=& n \, \eta \, \kappa_n/(Z\alpha) \,,\nonumber
\end{eqnarray}
and $E_{nlj}$ is the exact relativistic energy of the $nl_j$ state
for the Dirac-Coulomb problem.

The base integrals, defined as
\begin{equation}
\label{Kabc_int}
  K_{abc}(\kappa)=
  \int_0^1 dv \, \frac{v^{2a}}{(1-v^2)^{b/2}} \,
  \left( \frac{\kappa\sqrt{1-v^2}}{1+\kappa\sqrt{1-v^2}} \right)^c
  \,,
\end{equation}
are known in a closed form \cite{previous} (cf. \cite{CJP98}):
\[
  K_{abc}(\kappa)=
  \frac{1}{2}\kappa^c\,
  B\left(a+\frac12,1-\frac{b}{2}+\frac{c}{2}\right)
\]
\[
  \times
  {_3F_2}\left(\frac{c}{2},\, \frac{c}{2}+\frac12,\, 1-\frac{b}{2}+\frac{c}{2} ;\;
  \frac12,\, a+\frac32-\frac{b}{2}+\frac{c}{2} ;\; \kappa^2\right)
\]
\begin{equation}\label{defKabc}
  -\frac{c}{2}\,\kappa^{c+1}\,
  B\left(a+\frac12, \frac32-\frac{b}{2}+\frac{c}{2}\right)
\end{equation}
\[
  \times
  {_3F_2}\left(\frac{c}{2}+1,\, \frac{c}{2}+\frac12,\, \frac32-\frac{b}{2}+\frac{c}{2};\;
  \frac32,\, a+2-\frac{b}{2}+\frac{c}{2};\; \kappa^2\right)
\;.
\]
Here ${_3F_2}\bigl(\alpha,\beta,\gamma;\;\delta,\epsilon;\;
z\bigr)$ stands for the generalized hypergeometric function
\cite{gen3f2} and $B\bigl(\alpha,\beta\bigr)$ is the beta
function.

Such expressions  containing the hypergeometric function $_3F_2$
are cumbersome and far from being transparent; however, various
simpler asymptotic expressions are known
\cite{CJP98,soto,previous}. Here we intend to expand in $Z\alpha$.
The leading term of the $\Delta E(nl_j)$ expansion in $Z\alpha$
(see the $F$ term in (\ref{defFH})) was studied by us in detail
previously \cite{previous} (see also \cite{soto}). Now, we derive
from Eq.~(\ref{reldef}) the leading relativistic correction to
$\Delta E^{(0)}(nl_j)$, related to the $H$ term in (\ref{defFH}).

In the limit $Z\alpha\to0$ \eq{eurel} transforms to an expression
for the nonrelativistic correction to energy
\begin{eqnarray}
  \Delta E^{\rm (NR)}(nl) &=&
  -\frac{\alpha(Z\alpha)^2}{\pi n^2}\, (n+l)! \sum_{i,k=0}^{n_r} \frac{(-1)^{i+k}n_r!}{i!(n_r-i)!k!(n_r-k)!}
  \nonumber\\ &\times&
  \frac{(2l+i+k+1)!}{(2l+i+1)!(2l+k+1)!}
  \\ &\times&
  \left[
    K_{1,2,2l+i+k+2}(\kappa_n)
    -\frac{1}{3} K_{2,2,2l+i+k+2}(\kappa_n)
  \right]\;.
  \nonumber
\end{eqnarray}
This expression differs in form from those in \cite{soto} and
\cite{previous}, but agrees with them.

\section{Results for the low lying levels}

The explicit expression for the $H$ term is
\begin{equation}
\label{master_H}
H_{nlj}(\kappa_n)=
\frac{(n_r)!(j+n+1/2)!}{2(2j+1)(n-\nu)} \; \sum_{i,k=0}^{n_r}
\frac{(-1)^{i+k}}{i!(n_r-i)!k!(n_r-k)!}
\]
\[
\times
\frac{(2j+i+k)!}{(2j+i+1)!(2j+k+1)!}
\]
\[
\times
\Biggl\{
 \Biggl[
   2(i+k-2\nu)+\frac{\nu(2j+1)(4n-2j-i-k-1)}{n^2}
\]
\[
  +(2j+2i-2\nu+1)(2j+2k-2\nu+1)
\]
\[
\times
   \Biggl(
    \psi(2j+i+k+1) + \psi(j+n+3/2)
\]
\[
-\psi(2j+i+2)-\psi(2j+k+2)
    - \frac{(3n-2\nu)(2n-2j-1)}{4n^2(n-\nu)}
   \Biggr)
 \Biggr]
\]
\[
\times
  \left(
    K_{1,2,2j+i+k+1}(\kappa_n)
    -
    \frac{1}{3} K_{2,2,2j+i+k+1}(\kappa_n)
  \right)
\]
\[
+(2j+2i-2\nu+1)(2j+2k-2\nu+1)
\]
\[
  \times
  \Biggl[
    L_{1,2,2j+i+k+1}(\kappa_n)
    -
    \frac{1}{3} L_{2,2,2j+i+k+1}(\kappa_n)
\]
\[
-\frac{(2n-2j-1)(2j+i+k+1)}{4n^2\kappa_n}
\]
\[
\times \left(
    K_{1,3,2j+i+k+2}(\kappa_n)
    -
    \frac{1}{3} K_{2,3,2j+i+k+2}(\kappa_n)
  \right)
  \Biggr]
\Biggr\}
\;,
\end{equation}
where
\[
  L_{abc}(\kappa)=\frac{\partial K_{abc}(\kappa)}{\partial c}
  \,.
\]
Here $\psi(z)$ stands for the logarithmic derivative of the Euler
Gamma function
 \begin{eqnarray}
 \psi(n)&=& \psi(1) + \sum_{k=1}^{n-1}\frac{1}{k}\,,\nonumber\\
 \psi(n+1/2) &=& 2\psi(2n)-\psi(n)-2\ln2\nonumber\\
 &=& \psi(1)
 + 2\ln{2} + 2\sum_{k=0}^{n-1}\frac{1}{2k+1}\,,\nonumber
 \end{eqnarray}
and $-\psi(1)  = {\cal C}$ is the Euler constant, which finally
cancels out in (\ref{master_H}).

Various high-$\kappa$ asymptotics of $K_{abc}$ were studied
previously and applying (\ref{master_H}) to the low-lying states
we derive results for the relativistic correction $H_{nlj}$, which
are presented in Table~\ref{tabFS2} for states with $j=l+1/2$.
Separately we present in Table~\ref{tabFS} the asymptotic results
related to the difference $\Delta_{nl}^{\rm FS}$ defined in
Eq.~(\ref{deltafs}). The difference is the leading Uehling
contribution to the fine structure and thus it is the part of
$H_{nlj}(x)$ of most interest.

\begin{table}
\begin{center}
\caption{Asymptotics of the relativistic corrections $H_{nlj}(x)$
for some low-lying states. For $nl_j$ states $x=\kappa_n$.}
\label{tabFS2}
\begin{tabular}{cl}
\hline\noalign{\smallskip}
$nl_j$ & \hfil $H_{nlj}(x)$ \hfil
\\
\noalign{\smallskip}\hline\noalign{\smallskip}
 $1s_{1/2}$
 & $-\frac{1}{3}\ln(2x) - \frac{\pi^2}{9} + \frac{23}{18} - \frac{1}{2x^2}$
\\
\noalign{\smallskip}\hline\noalign{\smallskip}
 $2s_{1/2}$
 & $-\frac{5}{12}\ln(2x) - \frac{\pi^2}{9} + \frac{103}{72} - \frac{35}{16x^2}$
\\
 $2p_{1/2}$
 & $-\frac{5}{12}\ln(2x) - \frac{\pi^2}{9} + \frac{361}{216} - \frac{35}{16x^2}$
\\
\noalign{\smallskip}\hline\noalign{\smallskip}
 $3s_{1/2}$
 & $-\frac{1}{3}\ln(2x) - \frac{\pi^2}{9} + \frac{25}{18} - \frac{67}{18x^2}$
\\
 $3p_{1/2}$
 & $-\frac{1}{3}\ln(2x)
 - \frac{\pi^2}{9} + \frac{335}{216} - \frac{67}{18x^2}$
\\
 $3d_{3/2}$
 & $-\frac{1}{9}\ln(2x) - \frac{\pi^2}{18} + \frac{8279}{10800} - \frac{25}{18x^2}$
\\
\noalign{\smallskip}\hline\noalign{\smallskip}
 $4s_{1/2}$
 & $-\frac{13}{48}\ln(2x) -
 \frac{\pi^2}{9} + \frac{2327}{1728} - \frac{335}{64x^2}$
\\
 $4p_{1/2}$
 & $-\frac{13}{48}\ln(2x) - \frac{\pi^2}{9} + \frac{63287}{43200} - \frac{335}{64x^2}$
\\
 $4d_{3/2}$
 & $-\frac{5}{48}\ln(2x) - \frac{\pi^2}{18} + \frac{16243}{21600} - \frac{35}{16x^2}$
\\
 $4f_{5/2}$
 & $-\frac{7}{144}\ln(2x) - \frac{\pi^2}{27} +
 \frac{3038863}{6350400} - \frac{71}{64x^2}$
\\
\noalign{\smallskip}\hline
\end{tabular}
\end{center}
\end{table}

\begin{table}
\caption{Asymptotics of the Uehling contribution $\Delta_{nl}^{\rm
FS}(x)$ at $x\gg1$ for the fine structure of some of the lowest
states. For the $nl$ states $x=\kappa_n$.} \label{tabFS}
\begin{center}
\begin{tabular}{cl}
\hline\noalign{\smallskip}
$nl$ & \hfil $\Delta_{nl}^{\rm FS}(x)$ \hfil
\\
\noalign{\smallskip}\hline\noalign{\smallskip}
$2p$  & $\frac{1}{3}\ln{2x} +\frac{\pi^2}{18} -
\frac{215}{216} + \frac{27}{16x^2}$\\
\noalign{\smallskip}\hline\noalign{\smallskip}
$3p$ & $\frac{2}{9}\ln{2x} + \frac{\pi^2}{18} - \frac{355}{432} + \frac{7}{3x^2}$\\
$3d$ & $\frac{2}{27}\ln{2x} + \frac{\pi^2}{54} -
\frac{10559}{32400} + \frac{8}{9x^2}$\\
\noalign{\smallskip}\hline\noalign{\smallskip}
$4p$ & $\frac{1}{6}\ln{2x} + \frac{\pi^2}{18} - \frac{32101}{43200}
+ \frac{195}{64x^2}$\\
$4d$ & $\frac{1}{18}\ln{2x} + \frac{\pi^2}{54} -
\frac{36971}{129600} + \frac{69}{64x^2}$\\
$4f$ & $\frac{1}{36}\ln{2x} + \frac{\pi^2}{108} -
\frac{246373}{1587600} + \frac{39}{64x^2}$\\
\noalign{\smallskip}\hline
\end{tabular}
\end{center}
\end{table}

\section{Results for the near circular states}

The general expression (\ref{master_H}) contains a double
summation. However, for circular ($l=n-1$, $j=l+1/2$) and near
circular states only a few terms contribute to the sum and the result
is relatively simple. In the limit $\kappa_n\gg1$, the result for
small values of $n-l$ reads
\begin{equation}
 H_{n,n-1,n-1/2}(\kappa_n)
 =
 \frac{1}{n^2} \Biggl\{ \frac{1}{3}\biggl(-\ln(2\kappa_n)+
 \psi(2n)-\psi(1)\biggr)
\]
\[
 -\frac{2}{3}n\,\psi^{\prime}(2n)+\frac{5}{18}
 -\frac{1}{2}\frac{n^2}{\kappa_n^2} +{\cal O}\left(\frac{n^3}{\kappa_n^3}\right)\Biggr\}\,,
\end{equation}
\begin{equation}
 H_{n,n-1,n-3/2}(\kappa_n)
 =
\]
\[
 \frac{1}{n^2}\Biggl\{
 \frac{(n+3)}{3(n-1)}
 \biggl( -\ln(2\kappa_n)+\psi(2n)-\psi(1) \biggr)
\]
\[
 -\frac{2n^2}{3(n-1)}\psi^{\prime}(2n)
 +\frac{10n^3-21n^2-4n+12}{18(n-1)^2(2n-1)}
\]
\[
 -\frac{2n^3+6n^2-3n+1}{4n^2(n-1)}\,\frac{n^2}{\kappa_n^2}
 +{\cal O}\left(\frac{n^3}{\kappa_n^3}\right)\Biggr\}\,,
\end{equation}
\begin{equation}
 H_{n,n-2,n-3/2}(\kappa_n)=
\]
\[
 \frac{1}{n^2}\Biggl\{
 \frac{n+3}{3(n-1)}\biggl(-\ln(2\kappa_n)+\psi(2n)-\psi(1)\biggr)
 -\frac{2n^2}{3(n-1)}\psi^{\prime}(2n)
\]
\[
 +\frac{20n^4-76n^3+7n^2+64n-24}{18(n-1)^2(2n-1)^2}
\]
\[
 -\frac{2n^3+6n^2-3n+1}{4n^2(n-1)}\,\frac{n^2}{\kappa_n^2}
 +{\cal O}\left(\frac{n^3}{\kappa_n^3}\right)\Biggr\}\,,
\end{equation}
\begin{equation}
 H_{n,n-2,n-5/2}(\kappa_n)=
\]
\[
 \frac{1}{n^2}\Biggl\{
 \frac{n+6}{3(n-2)}
 \biggl( -\ln(2\kappa_n)+\psi(2n)-\psi(1)\biggr)
 -\frac{2n^2}{3(n-2)}\psi^{\prime}(2n)
\]
\[
 +\frac{40n^6-316n^5+510n^4+565n^3-1951n^2+1482n-342}{18(n-1)(n-2)^2(2n-1)^2(2n-3)}
\]
\[
 -\frac{n^3+6n^2-6n+4}{2n^2(n-2)}\,\frac{n^2}{\kappa_n^2}
 +{\cal O}\left(\frac{n^3}{\kappa_n^3}\right)\Biggr\}\,.
\end{equation}
The natural parameter of the expansion is $n/\kappa_n$, not just
$1/\kappa_n$, as one can observe in a limit $n\gg 1$
\begin{equation}
 H_{n,n-1,j}(\kappa_n) = H_{n,n-2,j}(\kappa_n)
 =
\]
\[
 \frac{1}{n^2}\Biggl\{
 \frac{1}{3}\left(\ln\left(\frac{n}{\kappa_n}\right)-\psi(1)\right)
 -\frac{1}{18}-\frac{1}{2}\frac{n^2}{\kappa_n^2}+\ldots\biggr\}\,,  \nonumber
\end{equation}
This parameter can be easily understood from the presentation in the
coordinate space (see \cite{previous} for more discussion).

For the fine-structure splitting we find
\begin{equation}
 \Delta^{\rm FS}_{n,n-1}=
 \frac{1}{n^3} \Biggl\{
 \frac{4}{3}\,\frac{n}{n-1}
 \biggl( \ln (2\kappa_n)-\psi(2n)+\psi(1)
 +\frac{n}{2}\,\psi^{\prime}(2n)\biggr)
\]
\[
 -\frac{n(4n^2-24n+17)}{18(n-1)^2(2n-1)}
 +\frac{8n^2-3n+1}{4n(n-1)}\,\frac{n^2}{\kappa_n^2}
\]
\[
 +{\cal O}\left(\frac{n^3}{\kappa_n^3}\right)\Biggr\}\,,
\end{equation}
\begin{equation}
 \Delta^{\rm FS}_{n,n-2}
 =
 \frac{1}{n^3} \Biggl\{
 \frac{4}{3}\,\frac{n^2}{(n-1)(n-2)}
\]
\[
 \times
 \biggl(
 \ln (2\kappa_n)-\psi(2n)+\psi(1)+\frac{n}{2}\,\psi^{\prime}(2n)
 \biggr)
\]
\[
 -\frac{n(16n^6-424n^5+1764n^4-2816n^3+1973n^2-576n+54)}{18(n-1)^2(n-2)^2(2n-1)^2(2n-3)}
\]
\[
 +\frac{8n^3-9n^2+13n-6}{4n(n-1)(n-2)}\,\frac{n^2}{\kappa_n^2}
 +{\cal O}\left(\frac{n^3}{\kappa_n^3}\right)\Biggr\}\,,
\end{equation}
and its asymptotics for $n\gg1$ are
\begin{equation}
  \Delta^{\rm FS}_{n,n-1}
  =
  \Delta^{\rm FS}_{n,n-2}
  =
\]
\[
  \frac{1}{n^3}\biggl\{
  \frac{4}{3}\left(\ln\left(\frac{\kappa_n}{n}\right)+\psi(1)\right)
  +\frac{2}{9}+\frac{2n^2}{\kappa_n^2}+\ldots\biggr\}\,.
\end{equation}

\section{Asymptotic behavior at low $\kappa_n$}

We are mostly interested in $\kappa_n \geq 1$ and $\kappa_n\gg 1$,
however, for the highly excited states one can arrive at the
situation when $\kappa$ is large, but $\kappa_n$ is small.

For the case of low $\kappa_n$ we obtain asymptotics of
\eq{master_H} for small $\kappa_n$
\begin{eqnarray}
  H_{nlj}(\kappa_n)
  &\approx&
  \kappa_n^{2j+1} \, \frac{(n+j+1/2)!(2j+3)!!}{2(2j+1)^3(n-\nu) (n_r)! (2j+1)!(2j+4)!!}
\nonumber\\
  &\times&
  \biggl\{
    -\frac{(2n-2j-1)^2\nu}{n^2}
  +(2j-2\nu+1)^2
\nonumber\\
  &\times&
  \biggl[
    \ln \kappa_n +\frac12 \psi(j+1/2)
    +\psi(j+n+3/2)
\nonumber\\
    &-&
    \psi(2j+2)
    -\frac12 \psi(j+3)
    -\frac{1}{2j+1}
    +\frac{1}{2j+3}
\nonumber\\
    &+&
    \frac{2j-2n+1}{4n^2}
    \left( 2j+3+\frac{n}{n-\nu}\right)
  \biggr]
  \biggr\}
  \,.
\end{eqnarray}

The above asymptotics prove to be especially simple for the case
$j<l$:

\begin{equation}
  H_{nlj}(\kappa_n)
  \approx
  - \kappa_n^{2j+1} \, \frac{(n+j+1/2)!(2j+3)!!}
  {n^2(2j+1)^3 (n_r-1)! (2j)!(2j+4)!!}
  \,.
\end{equation}

In the case of low-$\kappa$ another asymptotics can also be of
interest, namely a result in the leading order in $\kappa$ but
exact in $(Z\alpha)$:
\begin{equation}
  \Delta E(nl_j)
  \approx
  -\frac{\alpha}{\pi}
  \,
  \left( \frac{n\,\eta\,\kappa_n}{Z\alpha} \right)^{2\zeta}
  \,
  \frac{\zeta+1}{4\zeta^2(2\zeta+3)}
\]
\[
  \frac{\eta^2\Gamma(2\zeta+n_r+1)}{(n_r)!\left(\frac{Z\alpha}{\eta}-\nu\right)\Gamma(2\zeta)}
  \,
  B(3/2,\zeta)
  \times
\]
\[
  \times
  \left\{
  m\left[
    \left( \frac{Z\alpha}{\eta}-\nu \right)^2
    + n_r^2
  \right]
  - 2 n_r E_{nlj} \left( \frac{Z\alpha}{\eta}-\nu \right)
  \right\}
  \,.
\end{equation}


\section{Relativistic corrections in the logarithmic approximation}

To verify our calculations we consider a limit $\ln\kappa_n\gg 1$
and find the logarithmic terms within the effective charge
approach with the help of a substitute
\begin{equation}
  Z\alpha \longrightarrow
  Z\alpha(\kappa_n)
  =
  Z\alpha \left( 1+ \frac{2\alpha}{3\pi} \ln\kappa_n \right)
  \;.
\end{equation}
We find
\begin{eqnarray}\label{lamblog}
H^{\rm log}_{nlj}(\kappa_n)&=&-\frac{4}{3}
  \,  \frac{1}{n}
  \,  \left( \frac{1}{j+1/2} - \frac{3}{4n} \right)
  \,  \ln\kappa_n\;,
\nonumber\\
 \Delta^{\rm FS,\;log}_{nl}(\kappa_n)&=&
 \frac{4}{3}
  \,
  \frac{1}{n}
  \,
  \frac{1}{l(l+1)}
  \,
  \ln\kappa_n
 \;,
\end{eqnarray}
in agreement with the direct calculations above.

\section{Corrections due to the finite nuclear mass and anomalous
magnetic moment of the orbiting particle}

One feature of most exotic atoms is  that a gap between the
mass of the orbiting particle $m$ and of the nucleus $M$ is not so
large as in a conventional atom (where $m=m_e$ and thus $m/M\simeq
1/(1836\,A)$, where $A$ is the mass number). As a result, various
recoil effects become important. Another important property of
some exotic atoms is the presence of the anomalous magnetic moment
of the orbiting particle, which is not small. There are two kinds
of orbiting particles with spin 1/2 which can be described by the
Dirac equation. While the recoil effects are important for both
muonic and antiprotonic atoms, the effects of the non-zero
anomalous magnetic moment take place only in antiprotonic atoms
($\kappa'=(g_p-2)/2\simeq 1.79$)\footnote{The proton anomalous
magnetic moment is customarily denoted by $\kappa$. In our paper
(see, e.g., \cite{previous,CJP01}), we use $\kappa$ for the ratio
$(Z\alpha m)/m_e$ and here we denote the proton anomalous magnetic
moment by $\kappa'$.}.  The muon anomalous magnetic moment can be
treated perturbatively and taken into account together with
related radiative corrections.

We also note that the fine structure is considered as a structure
of levels due to the interaction of the spin of the orbiting particle
and the orbital moment. In this consideration we neglect the
nuclear magnetic moment and thus the hyperfine effects. If the
nucleus has a spin, the energy should be averaged over it to be
compared to our result. Such an asymmetric treatment is
reasonable when the nucleus is either spinless or heavier than
the orbiting particle and cannot be applied to, e.g., protonium, a
bound system of a proton and antiproton. In the case of heavy
nuclei it is customary to separate the fine and hyperfine
structure and for this case we consider below the recoil effects
in the fine-structure splitting.

Below we take into account the finite nuclear mass and anomalous
magnetic moment of the orbiting particles. To do that we re-visit
the well-known leading contributions to the fine structure, found
perturbatively, and show that their dependence on the nuclear mass
and the anomalous magnetic moment of the orbiting particle is of a
universal form. That allows us to adjust the results for
$\Delta^{\rm FS}$.

Above we calculated the leading contribution to the fine splitting
\begin{equation}
\Delta_{nl}^{\rm FS}(\kappa_n)
=H_{n,l,l+1/2}(\kappa_n)-H_{n,l,l-1/2}(\kappa_n)
\end{equation}
in the external field approximation and neglecting the muon
anomalous magnetic moment.

Now, we explain how to take into account the finite mass of the
nucleus and its anomalous magnetic moment for the fine structure.
Similar effects are also important for relativistic corrections in
general; however, the general case is much more complicated and
will be considered elsewhere.

The explicit results for $\Delta_{nl}^{\rm FS}(\kappa_n)$ were found
here by applying the Dirac equation. Let us consider this particular
effect perturbatively. In the case of $\kappa^\prime=0$ and $m/M=0$
we start from a Hamiltonian
\begin{equation}
H^{0} =H^{0}_{\rm NR}+H^{0}_{\rm FS}\;,
\end{equation}
where
 \begin{equation}\label{H0NRmR}
H^{0}_{\rm NR}= \frac{\mathbf{p}^2}{2m_R} + V(r)
\end{equation}
is the complete nonrelativistic Hamiltonian for the problem
$V(r)=V_C(r)+V_U(r)$, $m_R$ is the reduced mass, which in the
limit $m/M=0$ is equal to the particle mass $m$, and the
fine-structure effects are described by the potential
\begin{equation}
H^{0}_{\rm FS} = \frac{1}{2m^2}\,\frac{1}{r}\, \frac{\partial
V(r)}{\partial r} \, \left({\mbox{\boldmath $\rm s\cdot {\rm
l}$}}\right)\,.
\end{equation}

In terms of the eigenfunctions $\Psi^{0}_{\rm NR}$ of Hamiltonian
$H^{0}_{\rm NR}$, we present the fine structure as
\begin{equation}
h_{\rm FS} = \langle \Psi^{0}_{\rm NR}\vert H_{\rm FS}\vert
\Psi^{0}_{\rm NR}\rangle\,,
\end{equation}
where $h_{\rm FS}$ is a reduced Hamiltonian for variables related
to spin and angular momentum only. The complete fine structure,
which is not necessarily related to the pure Coulomb problem, is
eventually proportional to the matrix element of
\begin{equation}
\frac{1}{2m^2}\,\langle \Psi^{0}_{\rm NR}\vert \frac{1}{r}\,
\frac{\partial V(r)}{\partial r} \,  \vert \Psi^{0}_{\rm
NR}\rangle
\left({\mbox{\boldmath $\rm s\cdot {\rm l}$}}\right)\,.
\end{equation}

We note that if the potential $V(r)$ depends neither on the mass
of the orbiting particle $m$ nor on the nuclear mass $M$, the wave
function can depend only on the reduced mass $m_R$ because of the
kinetic energy and on $m_e$ in the particular case of the Uehling
potential.

Thence, from dimensional analysis we can find in the external field
approximation\footnote{A similar
result takes place for the whole fine-structure splitting and for
a contribution to it in any order in $\alpha$ once we consider a
perturbation of the Coulomb potential by the Uehling potential. In
our case, only a term linear in $\alpha$ is needed.}
\begin{equation}\label{Gdim}
\frac{\alpha}{\pi}\,(Z\alpha)^4 \, \frac{m
c^2}{n^2}\,\Delta_{nl}^{\rm FS}(\kappa_n) = m \, G(m/m_e)\;,
\end{equation}
where $G(x)$ is a dimensionless function, which depends on a state
$\Psi^{0}_{\rm NR}$.

What happens if we take into account the anomalous magnetic moment
and the finiteness of the nuclear mass? The nonrelativistic
Hamiltonian is now of the same form as before (see (\ref{H0NRmR})
) but the reduced mass
\[
m_R = \frac{Mm}{M+m}
\]
differs from the mass of the particle $m$. Meanwhile, the
Hamiltonian for the fine structure is of the form (cf., e.g.,
\cite{grotch71} for the pure Coulomb case)
\begin{eqnarray}\label{pureC}
H_{\rm FS}&=&\frac{1}{2m^2}\,\frac{1}{r}\,\frac{\partial
V(r)}{\partial r}\,\left({\mbox{\boldmath $\rm s\cdot {\rm
l}$}}\right)\cdot\left[(1+2\kappa')+
\frac{2m}{M}(1+\kappa')\right]
\nonumber\\
&=&
\frac{m_R^2}{m^2}\,\left[(1+2\kappa')+
\frac{2m}{M}(1+\kappa')\right]
\nonumber\\
&&
\phantom{mmmmmmmmmm}
\times
\frac{1}{r}\,\frac{1}{m_R^2}\frac{\partial
V(r)}{\partial r}\left({\mbox{\boldmath $\rm s\cdot {\rm
l}$}}\right)\,.
\end{eqnarray}
Thus, the result reads
\begin{equation}
m \Delta_{nl}^{\rm
FS}(\kappa_n)
\to
\frac{m_R^2}{m^2}\,\left[(1+2\kappa')+
\frac{2m}{M}(1+\kappa')\right]
\]
\[
\phantom{mmmmmmmm}
\times m_R\,\Delta_{nl}^{\rm
FS}(\kappa_R/n)
\end{equation}
and the splitting is
\begin{equation}
\frac{\alpha}{\pi}\,(Z\alpha)^4 \, \frac{m
c^2}{n^2}\,\frac{m_R^3}{m^3}\,\left[(1+2\kappa')+
\frac{2m}{M}(1+\kappa')\right]
\]
\[
\phantom{mmmmmmmmmm}
\times
\Delta_{nl}^{\rm
FS}(\kappa_R/n)\;.
\end{equation}
This result follows from dimentioanl analyses in (\ref{Gdim}) once
we take into account all $m$-factors in (\ref{pureC}). It includes
the complete mass dependence and the anomalous magnetic moment and
we can now use the results obtained above for the fine-structure
difference\footnote{We apply here the dimensional analysis only to
a term linear in perturbation $V_U$; however, the scaling factor,
which takes into account the nuclear mass and the anomalous
magnetic moment of the orbiting particle, is correct in any order
and for any nonrelativistic potential, as long as the latter does
not explicitly depend on $m$ or $M$.}
 $\Delta_{nl}^{\rm FS}(\kappa_n)$.

\section{Summary}

There is a wide variety of exotic atoms, which have been or may be
successfully produced; however, only in two kinds of such atoms
the orbiting particle has spin 1/2, namely in muonic and
antiprotonic atoms. In this paper we study fine structure caused
by the interaction of the spin of the orbiting particle and the
orbital moment. Effects of the nuclear spin and magnetic moment
are neglected. That is basically correct in two cases. Firstly,
one can study an exotic atom with a spinless nucleus. Secondly,
one can consider the energy averaged over the nuclear spin. The
latter is correct as an approximation and the corrections enter in
order $(m/M)^2$.

Concluding, we present here expressions for the fine splitting in
both muonic and antiprotonic atoms which consist of the leading
term and the Uehling correction and properly include the anomalous
magnetic moment and recoil effects.

Below we summarize results known in closed analytic form for
antiprotonic and muonic atoms. While the kinematic corrections
discussed in the previous section are well-known, in this paper we
found various presentations for the leading radiative correction
($\Delta_{nl}^{\rm FS}$) in Sects. 3--5.

\subsection{Antiprotonic atoms}

The QED result for the fine splitting in antiprotonic atoms is of
the form
\begin{eqnarray}
  \Delta E_{\rm FS}(nl)
  &=&
  \frac{1}{2}\,\frac{(Z\alpha)^4\, m_R\, c^2}{n^3}
  \,\left( \frac{m_R}{m} \right)^2\,
 \nonumber \\&\times&
  \left[ (1+2\kappa')+\frac{2m}{M}\,(1+\kappa') \right]
 \nonumber \\&\times&
  \left[\frac{1}{l(l+1)}+2n\,\frac{\alpha}{\pi}\,
\Delta_{nl}^{\rm FS} \left(\frac{\kappa_R}{n}\right)
  \right]
\end{eqnarray}
or
\begin{eqnarray}
 \Delta E_{\rm FS}(nl)
  &=&
  \frac{1}{2}\,\frac{(Z\alpha)^4\, m_R\, c^2}{n^3}
  \,\left( \frac{m_R}{m} \right)^2\,
 \nonumber \\&\times&
  \left[ (1+2\kappa')+\frac{2m}{M}\,(1+\kappa') \right]
 \nonumber \\&\times&
  \Biggl[\frac{1}{l(l+1)}
  \left(1+\frac{8}{3}\,\frac{\alpha}{\pi}\,\ln\left(\frac{\kappa_R}{n}\right)\right)
\nonumber\\
&&
\phantom{mmmm}
+2n\,\frac{\alpha}{\pi}\,
\Delta_{nl}^{\rm FS,\;c} \left(\frac{\kappa_R}{n}\right)
  \Biggr]\,,
\end{eqnarray}
where we introduced
\[
\Delta_{nl}^{\rm FS,\;c} (x)=\Delta_{nl}^{\rm FS}
(x)-\Delta_{nl}^{\rm FS,\;log} (x)\,,
\]
which does not contain the leading logarithmic term
(\ref{lamblog}) for high $\kappa_n$. The higher order corrections
include a small factor which is either $\alpha$ or $(Z\alpha)^2$.

The QED result above does not include any nucleus-related effects
such as the finite-size effects, virtual QED annihilation effects
and effects of strong interactions. The latter may be dominant for
low-$l$ states. On the contrary, for higher-$l$ states the
dominant correction to the fine structure is determined by the
equations above.

\subsection{Muonic atoms}

The QED result is of the form
\begin{eqnarray}
 \Delta E_{\rm FS}(nl)
  &=&\frac{1}{2}\,\frac{(Z\alpha)^4\, m_R\, c^2}{n^3}
  \,
  \left[ 1-\left( \frac{m_R}{M} \right)^2+\frac{\alpha}{\pi}\,\left(\frac{m_R}{m}\right)
  \right]
 \nonumber \\&\times&\left[\frac{1}{l(l+1)}+2n\,\frac{\alpha}{\pi}\,
\Delta_{nl}^{\rm FS} \left(\frac{\kappa_R}{n}\right)
  \right]
  \end{eqnarray}
or
\begin{eqnarray}
 \Delta E_{\rm FS}(nl)
  &=&\frac{1}{2}\,\frac{(Z\alpha)^4\, m_R\, c^2}{n^3}
  \,
  \left[ 1-\left( \frac{m_R}{M} \right)^2+\frac{\alpha}{\pi}\,\left(\frac{m_R}{m}\right)
  \right]
 \nonumber \\&\times&
\Biggl[\frac{1}{l(l+1)}\left(1+\frac{8}{3}\,\frac{\alpha}{\pi}\,\ln\left(\frac{\kappa_R}{n}\right)\right)
\nonumber\\&&
\phantom{mmmmmm}
+2n\,\frac{\alpha}{\pi}\,
\Delta_{nl}^{\rm FS,\;c} \left(\frac{\kappa_R}{n}\right)
  \Biggr]
 \,.
\end{eqnarray}

The higher order corrections include a small factor which is
either $\alpha$ or $(Z\alpha)^2$. Some of them have been known
only for a few particular cases, but not in general. The higher
order effects for some particular situations are reviewed, e.g.,
in \cite{borie,report}.

The difference with antiprotonic atoms is a perturbative treatment
of the anomalous magnetic moment of the orbiting particle. The
accuracy of the muonic expression is higher than in the
antiprotonic case, because of lack of the direct strong
interaction and smaller importance of the nuclear finite-size
effects because of a larger average distance between the orbiting
particle and the nucleus.

\section*{Acknowledgements}

This work was supported in part by RFBR (grant \# 06-02-16156) and
DFG (grant GZ 436 RUS 113/769/0-2).

\end{document}